\documentclass{article}
\usepackage{amsfonts,amssymb}
\usepackage[all, knot]{xy}
\xyoption{arc}
\input{epsf}

\newcommand{\Set}{{\rm Set}}
\newcommand{\Rel}{{\rm Rel}}
\newcommand{\Hilb}{{\rm Hilb}}
\newcommand{\Vect}{{\rm Vect}}
\newcommand{\Cob}{{\rm Cob}}
\newcommand{\C}{{\mathbb C}}
\newcommand{\M}{{\cal M}}

\renewcommand{\to}{\rightarrow}
\newcommand{\maps}{\colon}
\newcommand{\ten}{\otimes}
\newcommand{\tensor}{\otimes}
\newcommand{\iso}{\cong}

\begin{document}

      \begin{center}
      {\bf Quantum Quandaries: \\
       A Category-Theoretic Perspective \\ }
      \vspace{0.5cm}
      {\em John C.\ Baez\\}
      \vspace{0.3cm}
      {\small Department of Mathematics, University of California\\ 
      Riverside, California 92521 \\
      USA\\ }
      \vspace{0.3cm}
      {\small email: baez@math.ucr.edu\\}
      \vspace{0.3cm}
      {\small April 7, 2004 \\}
      \vspace{0.3cm}
{\small To appear in {\sl Structural Foundations of Quantum Gravity}, \\
eds.\ Steven French, Dean Rickles and Juha Saatsi, Oxford U.\ Press. }
      \end{center}

\begin{abstract}
\noindent
General relativity may seem very different from quantum theory, but 
work on quantum gravity has revealed a deep analogy between the two.
General relativity makes heavy use of the category $n\Cob$, whose 
objects are $(n-1)$-dimensional manifolds representing
`space' and whose morphisms are $n$-dimensional cobordisms representing 
`spacetime'.  Quantum theory makes heavy use of the category $\Hilb$, 
whose objects are Hilbert spaces used to describe `states', and whose 
morphisms are bounded linear operators used to describe `processes'.
Moreover, the categories $n\Cob$ and $\Hilb$ resemble each other far
more than either resembles $\Set$, the category whose objects are sets 
and whose morphisms are functions.  In particular, both $\Hilb$ and 
$n\Cob$ but not $\Set$ are $\ast$-categories with a noncartesian 
monoidal structure.  We show how this accounts for many of the famously
puzzling features of quantum theory: the failure of local realism, 
the impossibility of duplicating quantum information, and so on. 
We argue that these features only seem puzzling when we try to treat
$\Hilb$ as analogous to $\Set$ rather than $n\Cob$, so that
quantum theory will make more sense when regarded as part of 
a theory of spacetime.
\end{abstract}

\section{Introduction}

Faced with the great challenge of reconciling general relativity and quantum 
theory, it is difficult to know just how deeply we need to rethink basic 
concepts. By now it is almost a truism that the project of quantizing gravity 
may force us to modify our ideas about spacetime.  Could it also force us 
to modify our ideas about the quantum?  So far this thought has appealed 
mainly to those who feel uneasy about quantum theory and hope to replace 
it by something that makes more sense.  The problem is that the success 
and elegance of quantum theory make it hard to imagine promising
replacements.  Here I would like to propose another possibility, namely 
that {\it quantum theory will make more sense when regarded as part of 
a theory of spacetime}.  Furthermore, I claim that {\it we can only see 
this from a category-theoretic perspective} --- in particular, one that 
de-emphasizes the primary role of the category of sets and functions.   

Part of the difficulty of combining general relativity and quantum theory is
that they use different sorts of mathematics: one is based on objects 
such as manifolds, the other on objects such as Hilbert spaces.  As 
`sets equipped with extra structure', these look like very different 
things, so combining them in a single theory has always seemed a bit 
like trying to mix oil and water.  However, work on topological quantum 
field theory theory has uncovered a deep analogy between the two.  
Moreover, this analogy operates at the level of categories.  

We shall focus on two categories in this paper.  One is the category 
$\Hilb$ whose objects are Hilbert spaces and whose 
morphisms are linear operators between these.  This plays an important 
role in quantum theory.  The other is the category $n\Cob$ whose objects
are $(n-1)$-dimensional manifolds and whose morphisms are $n$-dimensional 
manifolds going between these.  This plays an important role in 
relativistic theories where spacetime is assumed to be $n$-dimensional:
in these theories the objects of $n\Cob$ represent possible choices of 
`space', while the morphisms --- called `cobordisms' --- represent 
possible choices of `spacetime'.

While an individual manifold is not very much like a Hilbert space,
the {\it category} $n\Cob$ turns out to have many structural
similarities to the {\it category} $\Hilb$.  The goal of this paper is
to explain these similarities and show that the most puzzling features
of quantum theory all arise from ways in which $\Hilb$ resembles
$n\Cob$ more than the category $\Set$, whose objects are sets and
whose morphisms are functions.

Since sets and functions capture many basic intuitions about macroscopic
objects, and the rules governing them have been incorporated into the 
foundations of mathematics, we naturally tend to focus on the fact that 
any quantum system has a {\it set} of states.  From a Hilbert space 
we can indeed extract a set of states, namely the set of unit vectors 
modulo phase.  However, this is often more misleading than productive,
because this process does not define a well-behaved map --- or more 
precisely, a functor --- from $\Hilb$ to $\Set$.  In some sense the gap 
between $\Hilb$ and $\Set$ is too great to be usefully bridged by this 
trick.  However, many of the ways in which $\Hilb$ differs from $\Set$ 
are ways in which it resembles $n\Cob$!  This suggests that the 
interpretation of quantum theory will become easier, not harder, when 
we finally succeed in merging it with general relativity.
  
\begin{figure}[ht]
\vskip 2em
\centerline{\epsfysize=1.2in\epsfbox{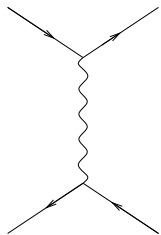}} 
\medskip
\caption{A Feynman diagram}
\label{feynman}
\end{figure}

In particular, it is easy to draw pictures of the objects and morphisms 
of $n\Cob$, at least for low $n$.  Doing so lets us {\it visualize} many
features of quantum theory.  This is not really a new discovery: it is 
implicit in the theory of Feynman diagrams.  Whenever one uses Feynman 
diagrams in quantum field theory, one is secretly working in some 
category where the morphisms are graphs with labelled edges and 
vertices, as shown in Figure \ref{feynman}.   

\noindent
The precise details of the category depend on the quantum field theory 
in question: the labels for edges correspond to the various 
{\it particles} of the theory, while the labels for vertices correspond 
to the {\it interactions} of the theory.  Regardless of the details, 
categories of this sort share many of the structural features of both 
$n\Cob$ and $\Hilb$.  Their resemblance to $n\Cob$, namely their 
topological nature, makes them a powerful tool for visualization.  On 
the other hand, their relation to $\Hilb$ makes them useful in 
calculations.
  
\begin{figure}[ht]
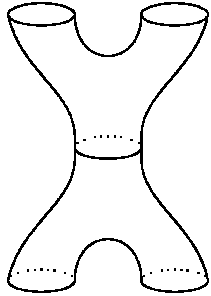

\vskip 2em
\[
  \xy 0 ;/r.35pc/:
  (0,0)*\ellipse(3,1){.};
  (0,0)*\ellipse(3,1)__,=:a(-180){-};
  (-3,-6)*\ellipse(3,1){.};
  (3,-6)*\ellipse(3,1){.};
  (-3,-6)*\ellipse(3,1)__,=:a(-180){-};
  (3,-6)*\ellipse(3,1)__,=:a(-180){-};
  (-3,-12)*{}="1";
  (3,-12)*{}="2";
  (-9,-12)*{}="A2";
  (9,-12)*{}="B2";
    "1";"2" **\crv{(-3,-7) & (3,-7)};
  (-3,0)*{}="A";
  (3,0)*{}="B";
  (-3,-1)*{}="A1";
  (3,-1)*{}="B1";
   "A";"A1" **\dir{-};
   "B";"B1" **\dir{-};
    "B2";"B1" **\crv{(8,-7) & (3,-5)};
    "A2";"A1" **\crv{(-8,-7) & (-3,-5)};
  (-3,6)*\ellipse(3,1){-};
  (3,6)*\ellipse(3,1){-};
  (-3,12)*{}="1";
  (3,12)*{}="2";
  (-9,12)*{}="A2";
  (9,12)*{}="B2";
    "1";"2" **\crv{(-3,7) & (3,7)};
  (-3,0)*{}="A";
  (3,0)*{}="B";
  (-3,1)*{}="A1";
  (3,1)*{}="B1";
   "A";"A1" **\dir{-};
   "B";"B1" **\dir{-};
    "B2";"B1" **\crv{(8,7) & (3,5)};
    "A2";"A1" **\crv{(-8,7) & (-3,5)};
 \endxy
\]
\medskip
\caption{A string worldsheet}
\label{string}
\end{figure}

Though Feynman diagrams are far from new, the fact that they are
morphisms in a category only became appreciated in work on quantum
gravity, especially string theory and loop quantum gravity.  Both
these approaches stretch the Feynman diagram concept in interesting
new directions.  In string theory, Feynman diagrams are replaced by
`string worldsheets': 2-dimensional cobordisms mapped into an ambient
spacetime, as shown in Figure \ref{string}.  Since these cobordisms no
longer have definite edges and vertices, there are no labels anymore.
This is one sense in which the various particles and interactions are
all unified in string theory.  The realization that processes in
string theory could be described as morphisms in a category was
crystallized by Segal's definition of `conformal field theory'
\cite{Segal}.  

\begin{figure}[ht]
\vskip 2em
\centerline{\epsfysize=1.5in\epsfbox{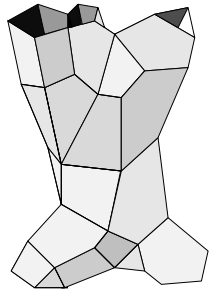}} 
\medskip
\caption{A spin foam}
\label{foam}
\end{figure}

Loop quantum gravity is moving towards a similar picture, though with 
some important differences.  In this approach processes are described 
by `spin foams'.  These are a 2-dimensional generalization of Feynman
diagrams built from vertices, edges and faces, as shown in Figure 
\ref{foam}.  They are not mapped into an ambient spacetime: in this 
approach spacetime is nothing but the spin foam itself --- or more 
precisely, a linear combination of spin foams.  Particles and 
interactions are not `unified' in these models, so there are labels on 
the vertices, edges and faces, which depend on the details of the 
model in question.  The category-theoretic underpinnings of spin foam 
models were explicit from the very beginning \cite{B2}, since they
were developed after Segal's work on conformal field theory, and also 
after Atiyah's work on topological quantum field theory \cite{Atiyah}, 
which exhibits the analogy between $n\Cob$ and $\Hilb$ in its simplest 
form.

There is not one whit of experimental evidence for either string
theory or loop quantum gravity, and both theories have some
serious problems, so it might seem premature for philosophers to 
consider their implications.  It
indeed makes little sense for philosophers to spend time chasing every
short-lived fad in these fast-moving subjects.  Instead, what is
worthy of reflection is that these two approaches to quantum gravity,
while disagreeing heatedly on so many issues \cite{Smolin,Vaas}, have
so much in common.  It suggests that in our attempts to reconcile the
quantum-theoretic notions of {\it state} and {\it process} with the
relativistic notions of {\it space} and {\it spacetime}, we have a
limited supply of promising ideas.  It is an open question whether
these ideas will be up to the task of describing nature.  But this
actually makes it more urgent, not less, for philosophers to clarify
and question these ideas and the implicit assumptions upon which they
rest.

Before plunging ahead, let us briefly sketch the contents of this
paper.  In Section \ref{TQFT} we explain the analogy between $n\Cob$
and $\Hilb$ by recalling Atiyah's definition of `topological quantum
field theory', or `TQFT' for short.  In Section \ref{*category}, we
begin by noting that unlike many familiar categories, neither $\Hilb$
nor $n\Cob$ is best regarded as a category whose objects are sets
equipped with extra structures and properties, and whose morphisms are
functions preserving these extra structures.  In particular, operators
between Hilbert spaces are not required to preserve the inner product.
This raises the question of precisely what role the inner product
plays in the category $\Hilb$.  Of course the inner product is crucial
in quantum theory, since we use it to compute transition amplitudes
between states --- but how does it manifest itself mathematically in the
structure of $\Hilb$?  One answer is that it gives a way to `reverse'
an operator $T \maps H \to H'$, obtaining an operator 
$T^\ast \maps H' \to H$ called the `adjoint' of $T$ such that
\[   \langle T^\ast \phi,\psi \rangle = \langle \phi, T\psi \rangle \]
for all $\psi \in H$ and $\phi \in H'$.
This makes $\Hilb$ into something called a `$\ast$-category': a
category where there is a built-in way to reverse any process.  As we
shall see, it is easy to compute transition amplitudes using the
$\ast$-category structure of $\Hilb$.  The category $n\Cob$
is also a $\ast$-category, where the adjoint of a spacetime is
obtained simply by switching the roles of future and past.  On the
other hand, $\Set$ cannot be made into a $\ast$-category.  All this
suggests that both quantum theory and general relativity will be best
understood in terms of categories quite different from the category of
sets and functions.
 
In Section \ref{monoidal} we tackle some of the most puzzling
features of quantum theory, namely those concerning joint systems:
physical systems composed of two parts.  It is in the study of joint
systems that one sees the `failure of local realism' that worried
Einstein so terribly \cite{EPR}, and was brought into clearer focus by
Bell \cite{Bell}.  Here is also where one discovers that one `cannot
clone a quantum state' --- a result due to Wooters and Zurek \cite{WZ}
which serves as the basis of quantum cryptography.  As explained in
Section \ref{monoidal}, both these phenomena follow from the failure
of the tensor product to be `cartesian' in a certain sense made
precise by category theory.  In $\Set$, the usual product of sets 
is cartesian, and this encapsulates many of our usual intuitions
about ordered pairs, like our ability to pick out the components $a$
and $b$ of any pair $(a,b)$, and our ability to `duplicate' any
element $a$ to obtain a pair $(a,a)$.  The fact that we cannot do
these things in $\Hilb$ is responsible for the failure of local
realism and the impossibility of duplicating a quantum state.  Here
again the category $\Hilb$ resembles $n\Cob$ more than $\Set$.  Like
$\Hilb$, the category $n\Cob$ has a noncartesian tensor product, given
by the disjoint union of manifolds.  Some of the mystery surrounding
joint systems in quantum theory dissipates when one focuses on the
analogy to $n\Cob$ and stops trying to analogize the tensor product of
Hilbert spaces to the Cartesian product of sets.

This paper is best read as a followup to my paper `Higher-Dimensional
Algebra and Planck-Scale Physics' \cite{B3}, since it expands on
some of the ideas already on touched upon there.  

\section{Lessons from Topological Quantum Field Theory} \label{TQFT}

\begin{figure}[ht]
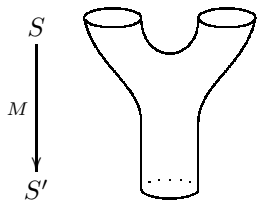

\vskip 2em
\[
  \xy 0 ;/r.30pc/:
  (0,-3)*\ellipse(3,1){.};
  (0,-3)*\ellipse(3,1)__,=:a(-180){-};
  (-3,6)*\ellipse(3,1){-};
  (3,6)*\ellipse(3,1){-};
  (-3,12)*{}="1";
  (3,12)*{}="2";
  (-9,12)*{}="A2";
  (9,12)*{}="B2";
    "1";"2" **\crv{(-3,7) & (3,7)};
  (-3,-6)*{}="A";
  (3,-6)*{}="B";
  (-3,1)*{}="A1";
  (3,1)*{}="B1";
   "A";"A1" **\dir{-};
   "B";"B1" **\dir{-};
    "B2";"B1" **\crv{(8,7) & (3,5)};
    "A2";"A1" **\crv{(-8,7) & (-3,5)};
    {\ar_{   M} (-14,11)*+{ S}; (-14,-6)*+{ S'}}
 \endxy
\]
\medskip
\caption{A cobordism}
\label{cobordism}
\end{figure}

Thanks to the influence of general relativity, there is a large
body of theoretical physics that does not presume a fixed topology 
for space or spacetime.  The idea is that after having assumed that
spacetime is $n$-dimensional, we are in principle free to choose any 
$(n-1)$-dimensional manifold to represent space at a given time.  
Moreover, given two such manifolds, say $S$ and $S'$, we are free to 
choose any $n$-dimensional manifold-with-boundary, say $M$, 
to represent the portion of spacetime between them, so long as 
the boundary of $M$ is the union of $S$ and $S'$.  In this situation 
we write $M \maps S \to S'$, even though $M$ is not a function from 
$S$ to $S'$, because we may think of $M$ as the process of time passing 
from the moment $S$ to the moment $S'$.  Mathematicians call $M$ a 
{\bf cobordism} from $S$ to $S'$.  For example, in Figure 
\ref{cobordism} we depict a 2-dimensional manifold $M$ going from 
a 1-dimensional manifold $S$ (a pair of circles) to a 1-dimensional 
manifold $S'$ (a single circle).   Physically, this cobordism 
represents a process in which two separate spaces collide to form a 
single one!  This is an example of `topology change'.

All this has a close analogue in quantum theory.  First, just as we can
use any $(n-1)$-manifold to represent space, we can use any 
Hilbert space to describe the states of some quantum system.  
Second, just as we can use any cobordism to represent
a spacetime going from one space to another, we can use any operator to 
describe a process taking states of one system to states of another.  
More precisely, given systems whose states are described using the 
Hilbert spaces $H$ and $H'$, respectively, any bounded linear operator 
$T \maps H \to H'$ describes a process that carries states of the first 
system to states of the second.  We are most comfortable with this idea
when the operator $T$ is unitary, or at least an isometry.
After all, given a state described as a unit vector $\psi \in H$, we can 
only be sure $T\psi$ is a unit vector in $H'$ if $T$ is an isometry.
So, only in this case does $T$ define a {\it function} from the {\it set}
of states of the first system to the {\it set} of states of the second.
However, the interpretation of linear operators as processes makes 
sense more generally.  One way to make this interpretation precise is 
as follows: given a unit vector $\psi \in H$ and an orthonormal basis 
$\phi_i$ of $H'$, we declare that the {\bf relative probability} for a 
system prepared in the state $\psi$ to be observed in the state $\phi_i$ 
after undergoing the process $T$ is $|\langle \phi_i, T\psi \rangle|^2$.  
By this, we mean that the probability of observing the system in the 
$i$th state divided by the probability of observing it in the $j$th 
state is
\[
     \frac{|\langle \phi_i, T\psi \rangle|^2}
          {|\langle \phi_j, T\psi \rangle|^2} .
\]

The use of nonunitary operators to describe quantum processes is
not new.  For example, projection operators have long been used to
describe processes like sending a photon through a polarizing filter. 
However, these examples traditionally arise when we treat
part of the system (e.g.\ the measuring apparatus) classically.  It is
often assumed that at a fundamental level, the laws of nature in 
quantum theory describe time evolution using unitary operators.  But 
as we shall see in Section \ref{*category}, this assumption should 
be dropped in theories where the topology of space can change.  In such 
theories we should let {\it all} the morphisms in $\Hilb$ qualify as 
`processes', just as we let {\it all} morphisms in $n\Cob$ qualify as 
spacetimes. 
 
Having clarified this delicate point, we are now in a position to 
clearly see a structural analogy between general relativity
and quantum theory, in which $(n-1)$-dimensional manifolds representing 
space are analogous to Hilbert spaces, while cobordisms describing 
spacetime are analogous to operators.  Indulging in some lofty 
rhetoric, we might say that {\it space} and {\it state} are aspects of 
{\it being}, while {\it spacetime} and {\it process} are aspects of 
{\it becoming}.  We summarize this analogy in Table 1.

\vskip 2em
\begin{center}
\small
\begin{tabular}{|c|c|}                    \hline
GENERAL RELATIVITY                &  QUANTUM THEORY      \\  \hline
$(n-1)$-dimensional manifold      &  Hilbert space       \\  
(space)                           & (states)             \\  \hline
cobordism between $(n-1)$-dimensional manifolds &  operator between 
                                      Hilbert spaces     \\ 
(spacetime)                       &  (process)           \\  \hline
composition of cobordisms         &  composition of operators \\  \hline
identity cobordism                &  identity operator        \\  \hline
\end{tabular} \vskip 1em
Table 1: Analogy between general relativity and quantum theory
\end{center}
\vskip 0.5em

This analogy becomes more than mere rhetoric when applied to
topological quantum field theory \cite{B3}.   
In quantum field theory on curved spacetime, space and spacetime are not 
just manifolds: they come with fixed `background metrics' that allow us 
to measure distances and times.   In this context, $S$ and $S'$ are 
Riemannian manifolds, and $M \maps S \to S'$ is a {\bf Lorentzian 
cobordism} from $S$ to $S'$: that is, a Lorentzian manifold with 
boundary whose metric restricts at the boundary to the metrics on $S$ 
and $S'$.  However, topological quantum field theories are an attempt 
to do background-free physics, so in this context we drop the background 
metrics: we merely assume that space is an $(n-1)$-dimensional manifold 
and spacetime is a cobordism between such manifolds.  A topological
quantum field theory then consists of a map $Z$ assigning a Hilbert 
space of states $Z(S)$ to any $(n-1)$-manifold $S$ and a linear operator 
$Z(M) \maps Z(S) \to Z(S')$ to any cobordism between such manifolds.  
This map cannot be arbitrary, though: for starters, it must be a 
{\it functor} from the category of $n$-dimensional cobordisms to the 
category of Hilbert spaces.  This is a great example of how every 
sufficiently good analogy is yearning to become a functor.  

However, we are getting a bit ahead of ourselves.  Before we can
talk about functors, we must talk about categories.  What is the 
category of $n$-dimensional cobordisms, and what is the category of 
Hilbert spaces?  The answers to these questions will allow us to make 
the analogy in Table 1 much more precise.

First, recall that a {\bf category} consists of a collection of 
objects, a collection of morphisms $f \maps A \to B$ from any object 
$A$ to any object $B$, a rule for composing morphisms $f \maps A \to B$ 
and $g \maps B \to C$ to obtain a morphism $gf \maps A \to C$, and for 
each object $A$ an identity morphism $1_A \maps A \to A$.  These must 
satisfy the associative law $f(gh) = (fg)h$ and the left and right unit 
laws $1_A f = f$ and $f 1_A = f$ whenever these composites are defined.
In many cases, the objects of a category are best thought of as
{\it sets equipped with extra structure}, while the morphisms are 
{\it functions preserving the extra structure}.  However, this is true
neither for the category of Hilbert spaces nor for the category of 
cobordisms.  

In the category $\Hilb$ we take the objects to be Hilbert spaces
and the morphisms to be bounded linear operators.  Composition and 
identity operators are defined as usual.  Hilbert spaces are indeed 
sets equipped with extra structure, but bounded linear operators do 
not preserve all this extra structure: in particular, they need not 
preserve the inner product.  This may seem like a fine point, but it 
is important, and we shall explore its significance in detail in 
Section \ref{*category}.

\begin{figure}[t]
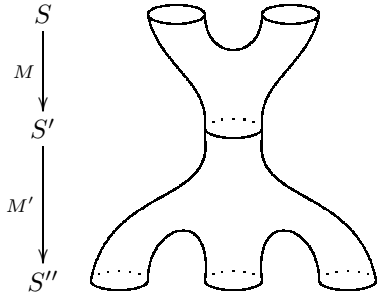

\vskip 2em
\[
  \xy 0 ;/r.30pc/:
    (0,0)*\ellipse(3,1){.};
    (0,0)*\ellipse(3,1)__,=:a(-180){-};
  (-6,-8)*\ellipse(3,1){.};
  (6,-8)*\ellipse(3,1){.};
  (0,-8)*\ellipse(3,1){.};
  (-6,-8)*\ellipse(3,1)__,=:a(-180){-};
  (6,-8)*\ellipse(3,1)__,=:a(-180){-};
  (0,-8)*\ellipse(3,1)__,=:a(180){-};
 (-3,6)*\ellipse(3,1){-};
  (3,6)*\ellipse(3,1){-};
  (-3,12)*{}="1";
  (3,12)*{}="2";
  (-9,12)*{}="A2";
  (9,12)*{}="B2";
    "1";"2" **\crv{(-3,7) & (3,7)};
  (-3,0)*{}="A";
  (3,0)*{}="B";
  (-3,1)*{}="A1";
  (3,1)*{}="B1";
   "A";"A1" **\dir{-};
   "B";"B1" **\dir{-};
    "B2";"B1" **\crv{(8,7) & (3,5)};
    "A2";"A1" **\crv{(-8,7) & (-3,5)};
  (3,-16)*{}="1";
  (9,-16)*{}="2";
      "1";"2" **\crv{(2,-8) & (9,-10)};
  (-3,-16)*{}="1";
  (-9,-16)*{}="2";
      "1";"2" **\crv{(-2,-8) & (-9,-10)};
  (-15,-16)*{}="A2";
  (15,-16)*{}="B2";
  (-3,0)*{}="A";
  (3,0)*{}="B";
  (-3,-1)*{}="A1";
  (3,-1)*{}="B1";
   "A";"A1" **\dir{-};
   "B";"B1" **\dir{-};
    "B2";"B1" **\crv{(13,-6) & (2,-8)};
    "A2";"A1" **\crv{(-13,-6) & (-2,-8)};
        {\ar_{M} (-20,12)*+{S}; (-20,0)*+{S'}="1"};
        {\ar_{M'} "1"; (-20,-16)*+{S''}};
 \endxy
\]
\medskip
\caption{Composition of cobordisms}
\label{composition}
\end{figure}

In the category $n\Cob$ we take the objects to be $(n-1)$-dimensional
manifolds and the morphisms to be cobordisms between these.  (For 
technical reasons mathematicians usually assume both to be compact and 
oriented.)  Here the morphisms are not functions at all!  Nonetheless 
we can `compose' two cobordisms $M\maps S \to S'$ and 
$M' \maps S' \to S''$, obtaining a cobordism $M' M \maps S \to S''$, as
in Figure \ref{composition}.  The idea here is that the passage of time 
corresponding to $M$ followed by the passage of time corresponding to 
$M'$ equals the passage of time corresponding to $M'M$.   This is 
analogous to the familiar idea that waiting $t$ seconds followed by 
waiting $t'$ seconds is the same as waiting $t'+t$ seconds. 
The big difference is that in topological quantum field theory we 
cannot measure time in seconds, because there is no background metric 
available to let us count the passage of time.  We can only keep track 
of topology change.  Just as ordinary addition is associative, 
composition of cobordisms satisfies the associative law:
\[                (M'' M') M = M'' (M' M)     .        \] 

\begin{figure}[h]
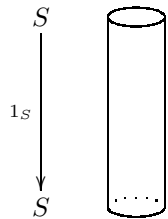

\vskip 2em
\[
  \xy 0 ;/r.30pc/:
  (0,-5)*\ellipse(3,1){.};
  (0,-5)*\ellipse(3,1)__,=:a(-180){-};
  (0,5)*\ellipse(3,1){-};
  (-3,10)*{}="1";
  (3,10)*{}="2";
  (-3,-10)*{}="1'";
  (3,-10)*{}="2'";
   "1"; "1'" **\dir{-};
   "2"; "2'" **\dir{-};
    {\ar_{1_S} (-10,10)*+{S}; (-10,-10)*+{S}};
 \endxy
\]
\medskip
\caption{An identity cobordism}
\label{identity}
\end{figure}

\noindent
Furthermore, for any $(n-1)$-dimensional manifold $S$ representing space,
there is a cobordism $1_S \maps S \to S$ called the `identity'
cobordism, which represents a passage of time during which the topology
of space stays constant.  For example, when $S$ is a circle, the identity 
cobordism $1_S$ is a cylinder, as shown in Figure \ref{identity}.   In 
general, the identity cobordism $1_S$ has the property that 
\[                         1_S M = M  \]
and
\[                         M 1_S  = M \]
whenever these composites are defined.
These properties say that an identity cobordism is analogous to waiting 
0 seconds: if you wait 0 seconds and then wait $t$ more seconds, or 
wait $t$ seconds and then wait 0 more seconds, this is the same as 
waiting $t$ seconds.

A {\bf functor} between categories is a map sending objects to objects
and morphisms to morphisms, preserving composition and identities.  Thus, 
in saying that a topological quantum field theory is a functor
\[            Z \maps n\Cob \to \Hilb  , \]
we merely mean that it assigns a Hilbert space of states $Z(S)$ to 
any $(n-1)$-dimensional manifold $S$ and a linear operator 
$Z(M) \maps Z(S) \to Z(S')$ to any $n$-dimensional cobordism 
$M \maps S \to S'$ in such a way that:
\begin{itemize}
\item For any $n$-dimensional cobordisms 
$M \maps S \to S'$ and $M' \maps S' \to S''$, 
\[                Z(M'M) = Z(M')Z(M) .  \]
\item For any $(n-1)$-dimensional manifold $S$, 
\[                Z(1_S) = 1_{Z(S)}  .\] 
\end{itemize} 
Both these axioms make sense if one ponders them a bit. 
The first says that the passage of time corresponding to the cobordism 
$M$ followed by the passage of time corresponding to $M'$ has the  same
effect on a state as the combined passage of time corresponding to
$M'M$.  The second says that a passage of time in which no topology 
change occurs has no effect at all on the state of the universe.  This
seems paradoxical at first, since it seems we regularly observe things 
happening even in the absence of topology change.   However, this
paradox is easily resolved: a topological quantum field theory describes 
a world without local degrees of freedom.   In 
such a world, nothing local happens, so the state of the universe can 
only change when the topology of space itself changes.  

Unless elementary particles are wormhole ends or some other sort
of topological phenomenon, it seems our own world is quite unlike 
this.  Thus, we hasten to reassure the 
reader that this peculiarity of topological quantum field theory is 
{\it not} crucial to our overall point, which is the analogy between 
categories describing space and spacetime and those describing quantum 
states and processes.  If we were doing quantum field theory on curved 
spacetime, we would replace $n\Cob$ with a category where the objects 
are $n$-dimensional Riemannian manifolds and most of the morphisms are 
Lorentzian cobordisms between these.  In this case a cobordism 
$M \maps S \to S'$ has not just a topology but also a geometry, so we 
can use cylinder-shaped cobordisms of different `lengths' to describe 
time evolution for different amounts of time.  The identity morphism 
is then described by a cylinder of `length zero'.  This degenerate
cylinder is not really a Lorentzian cobordism, which leads to some 
technical complications.  However, Segal showed how to get around these
in his axioms for a conformal field theory \cite{Segal}.  There are 
some further technical complications arising from the fact that except 
in low dimensions, we need to use the C*-algebraic approach to quantum 
theory, instead of the Hilbert space approach \cite{Earman}.  Here the
category $\Hilb$ should be replaced by one where the objects are 
C$^*$-algebras and the morphisms are completely positive maps between
their duals \cite{HMS}.

Setting aside these nuances, our main point is that treating a TQFT as
a functor from $n\Cob$ to $\Hilb$ is a way of making very precise some
of the analogies between general relativity and quantum theory.
However, we can go further!  A TQFT is more than just a functor.  It
must also be compatible with the `monoidal category' structure of
$n\Cob$ and $\Hilb$, and to be physically well-behaved it must also be
compatible with their `$\ast$-category' structure.  We examine these
extra structures in the next two sections.

\section{The $\ast$-Category of Hilbert Spaces}  
\label{*category}

What is the category of Hilbert spaces?  While we have already given
an answer, this is actually a tricky question, one that makes many
category theorists acutely uncomfortable.  

To understand this, we must start by recalling that one use of 
categories is to organize discourse about various sorts of 
`mathematical objects': groups, rings, vector spaces, topological 
spaces, manifolds and so on.  Quite commonly these mathematical objects 
are sets equipped with extra structure and properties, so let us 
restrict attention to this case.   Here by {\bf structure} we mean 
operations and relations defined on the set in question, while by 
{\bf properties} we mean axioms that these operations and relations 
are required to satisfy.  The division into structure and properties 
is evident from the standard form of mathematical definitions such as 
``a widget is a set equipped with ... such that ....''   Here the 
structures are listed in the first blank, while the properties are 
listed in the second.    

To build a category of this sort of mathematical object, we must also 
define morphisms between these objects.  When the objects are {\it 
sets equipped with extra structure and properties}, the morphisms are 
typically taken to be {\it functions that preserve the extra structure}.
At the expense of a long digression we could make this completely 
precise --- and also more general, since we can also build categories 
by equipping not sets but objects of other categories with extra 
structure and properties.  However, we prefer to illustrate the idea 
with an example.  We take an example closely related to but subtly 
different from the category of Hilbert spaces: the category of 
complex vector spaces.  

A {\bf complex vector space} is a set $V$ equipped with 
extra structure consisting of operations called addition
\[              + \maps V \times V \to V  \]
and scalar multiplication
\[              \cdot \maps \C \times V \to V ,\]
which in turn must have certain extra properties: commutativity
and associativity together with the existence of an 
identity and inverses for addition, associativity and the unit law for 
scalar multiplication, and distributivity of scalar multiplication
over adddition.   Given vector spaces $V$ and $V'$, a
{\bf linear operator} $T \maps V \to V'$ can be defined as 
a function preserving all the extra structure.  This means
that we require 
\[     T(\psi + \phi) = T(\psi) + T(\phi) \]
and 
\[        T(c\psi) = c T(\psi)  \]
for all $\psi,\phi \in V$ and $c \in \C$.  Note that the 
properties do not enter here.  Mathematicians
define the category $\Vect$ to have complex vector spaces 
as its objects and linear operators between them as its
morphisms.

Now compare the case of Hilbert spaces.  A Hilbert space $H$
is a set equipped with all the structure of a complex vector
space but also some more, namely an inner product
\[        \langle \cdot, \cdot \rangle \maps H \times H \to \C . \]
Similarly, it has all the properties of a complex vector spaces
but also some more: for all $\phi, \psi, \psi' \in H$
and $c \in \C$ we have the equations
\[    \langle \phi , \psi + \psi' \rangle = 
       \langle \phi, \psi \rangle + \langle \phi, \psi' \rangle, \]
\[   \langle \phi, c\psi \rangle = c \langle \phi, \psi \rangle, \]
\[   \langle \phi,\psi \rangle = \overline{\langle \psi,\phi \rangle},\]
together with the inequality
\[   \langle \psi , \psi \rangle \ge 0 \]
where equality holds only if $\psi = 0$; furthermore, the norm defined
by the inner product must be complete.  
Given Hilbert spaces $H$ and $H'$, a function $T \maps H \to H'$
that preserves all the structure is thus a linear operator that 
preserves the inner product:
\[      \langle T \phi, T \psi \rangle = \langle \phi,\psi \rangle \]
for all $\phi,\psi \in H$.  Such an operator is called an 
{\bf isometry}.    

If we followed the pattern that works for vector spaces 
and many other mathematical objects, we would thus define the
category $\Hilb$ to have Hilbert spaces as objects and 
isometries as morphisms.  However, this category seems
too constricted to suit what physicists actually do
with Hilbert spaces: they frequently need operators
that aren't isometries!   Unitary operators are always 
isometries, but self-adjoint operators, for example, are not.

The alternative we adopt in this paper is to work with the category
$\Hilb$ whose objects are Hilbert spaces and the morphisms are bounded
linear operators.  However, this leads to a curious puzzle.  In a 
precise technical sense, the category of finite-dimensional Hilbert 
spaces and linear operators between these is {\it equivalent} to the 
category of finite-dimensional complex vector spaces and linear operators.  
So, in defining this category, we might as well ignore the inner product
entirely!  The puzzle is thus what role, if any, the inner product
plays in this category.

The case of general, possibly infinite-dimensional Hilbert spaces
is subtler, but the puzzle persists.  The category of all Hilbert
spaces and bounded linear operators between them is {\it not}
equivalent to the category of all complex vector spaces and linear
operators.  However, it {\it is} equivalent to the category 
of `Hilbertizable' vector spaces --- that is, vector spaces equipped
with a topology coming from {\it some} Hilbert space structure ---
and continuous linear operators between these.  So, in defining
this category, what matters is not the inner product but merely
the topology it gives rise to.  The point is that bounded linear operators
don't preserve the inner product, just the topology, and a structure
that is not preserved might as well be ignored, as far as the category
is concerned.

My resolution of this puzzle is simple but a bit upsetting to
most category theorists.  I admit that the inner product is inessential
in defining the category of Hilbert spaces and bounded linear
operators.  However, I insist that it plays a crucial role in making 
this category into a $\ast$-category!

What is a $\ast$-category?  It is a category $C$
equipped with a map sending each morphism $f \maps X \to Y$ to
a morphism $f^\ast \maps Y \to X$, satisfying
\[             1_X^\ast = 1_X ,\]
\[            (fg)^\ast = g^\ast f^\ast, \]
and 
\[          f^{\ast\ast} = f. \]
To make $\Hilb$ into a $\ast$-category we
define $T^\ast$ for any bounded linear operator $T \maps H \to H'$
to be the {\bf adjoint} operator $T^\ast H' \to H$, given
by 
\[   \langle T^\ast \psi,\phi \rangle = \langle \psi, T\phi \rangle .\]
We see by this formula that the inner product on both $H$ and $H'$
are required to define the adjoint of $T$.

In fact, we can completely recover the inner product on every
Hilbert space from the $\ast$-category structure of $\Hilb$.  
Given a Hilbert space $H$ and a vector $\psi \in H$,
there is a unique operator $T_\psi \maps \C \to H$ with $T_\psi(1) 
= \psi$.  Conversely, any operator from $\C$ to $H$ determines a
unique vector in $H$ this way.  So, we can think of elements of a 
Hilbert space as morphisms from $\C$ to this Hilbert space.  Using 
this trick, an easy calculation shows that
\[        \langle \phi,\psi \rangle = {T_\phi}^{\!\!\ast}\, T_\psi . \]
The right-hand side is really a linear operator from $\C$ to
$\C$, but there is a canonical way to identify such a thing with a
complex number.  So, everything about inner products is encoded in
the $\ast$-category structure of $\Hilb$.  Moreover, this way of
thinking about the inner product formalizes an old idea of Dirac.
The operator $T_\psi$ is really just Dirac's `ket' $|\psi\rangle$,
while ${T_\phi}^{\!\!\ast}$ is the `bra' $\langle \phi |$.  Composing
a ket with a bra, we get the inner product.

This shows how adjoints are closely tied to the inner product
structure on Hilbert space.  But what is the physical significance 
of the adjoint of an operator, or more generally the $\ast$ operation 
in any $\ast$-category?   Most fundamentally, the $\ast$ operation
gives us a way to `reverse' a morphism even when it is not 
invertible.  If we think of inner products as giving transition 
amplitudes between states in quantum theory, the equation
$ \langle T^\ast \phi,\psi \rangle = \langle \phi, T\psi \rangle $
says that $T^\ast$ is the unique operation we can perform on any state
$\phi$ so that the transition amplitude from $T \psi$ to $\phi$
is the same as that from $\psi$ to $T^\ast \phi$. 
So, in a suggestive but loose way, we can say that the process 
described by $T^\ast$ is some sort of `time-reversed' version of the 
process described by $T$.  If $T$ is unitary, $T^\ast$ is just the 
inverse of $T$.  But, $T^\ast$ makes sense even when $T$ has no inverse!  

\begin{figure}[h]
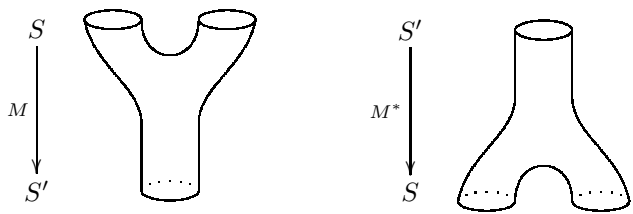

\vskip 2em
\[
 \vcenter{
  \xy 0 ;/r.30pc/:
  (0,-3)*\ellipse(3,1){.};
  (0,-3)*\ellipse(3,1)__,=:a(-180){-};
  (-3,6)*\ellipse(3,1){-};
  (3,6)*\ellipse(3,1){-};
  (-3,12)*{}="1";
  (3,12)*{}="2";
  (-9,12)*{}="A2";
  (9,12)*{}="B2";
    "1";"2" **\crv{(-3,7) & (3,7)};
  (-3,-6)*{}="A";
  (3,-6)*{}="B";
  (-3,1)*{}="A1";
  (3,1)*{}="B1";
   "A";"A1" **\dir{-};
   "B";"B1" **\dir{-};
    "B2";"B1" **\crv{(8,7) & (3,5)};
    "A2";"A1" **\crv{(-8,7) & (-3,5)};
    {\ar_{M} (-14,11)*+{S}; (-14,-6)*+{S'}}
 \endxy}
\qquad  \qquad
 \vcenter{
  \xy 0 ;/r.30pc/:
  (0,3)*\ellipse(3,1){-};
  (-3,-6)*\ellipse(3,1){.};
  (-3,-6)*\ellipse(3,1)__,=:a(-180){-};
  (3,-6)*\ellipse(3,1){.};
  (3,-6)*\ellipse(3,1)__,=:a(-180){-};
  (-3,-12)*{}="1";
  (3,-12)*{}="2";
  (-9,-12)*{}="A2";
  (9,-12)*{}="B2";
    "1";"2" **\crv{(-3,-7) & (3,-7)};
  (-3,6)*{}="A";
  (3,6)*{}="B";
  (-3,-1)*{}="A1";
  (3,-1)*{}="B1";
   "A";"A1" **\dir{-};
   "B";"B1" **\dir{-};
    "B2";"B1" **\crv{(8,-7) & (3,-5)};
    "A2";"A1" **\crv{(-8,-7) & (-3,-5)};
    {\ar_{M^{\ast}} (-14,6)*+{S'}; (-14,-11)*+{S}}
 \endxy}
\]
\medskip
\caption{A cobordism and its adjoint}
\label{adjoint}
\end{figure}

This suggestive but loose relation between $\ast$ operations and 
time reversal becomes more precise in the case of $n\Cob$.
Here the $\ast$ operation really {\it is} time reversal.  More
precisely, given an $n$-dimensional cobordism $M \maps S \to S'$,
we let the {\bf adjoint} cobordism $M^\ast \maps S' \to S$ to be 
the same manifold, but with the `past' and `future' parts of its
boundary switched, as in Figure \ref{adjoint}.  It is easy to check
that this makes $n\Cob$ into a $\ast$-category.

In a so-called {\bf unitary} topological quantum field theory 
(the terminology is a bit unfortunate), we demand that the
functor $Z \maps n\Cob \to \Hilb$ preserve the $\ast$-category
structure in the following sense:
\[             Z(M^\ast) = Z(M)^\ast  .\]
All the TQFTs of interest in physics have this property, and a similar
property holds for conformal field theories and other quantum field
theories on curved spacetime.  This means that in the analogy between
general relativity and quantum theory, {\it the analogue of time
reversal is taking the adjoint of an operator between Hilbert spaces}.
To `reverse' a spacetime $M \maps S \to S'$ we formally switch the
notions of future and past, while to `reverse' a process $T \maps H
\to H'$ we take its adjoint.

Taking this analogy seriously leads us in some interesting directions.
First, since the $\ast$ operation in $n\Cob$ is given by time
reversal, while $\ast$ operation in $\Hilb$ is defined using the inner
product, there should be some relation between time reversal and the
inner product in quantum theory!  The details remain obscure, at least
to me, but we can make a little progress by pondering the following
equation, which we originally introduced as a `trick' for expressing
inner products in terms of adjoint operators:
\[        \langle \phi,\psi \rangle = {T_\phi}^{\!\!\ast} \, T_\psi . \]
An equation this important should not be a mere trick!  To try to
interpret it, suppose that in some sense the operator $T_\psi$
describes `the process of preparing the system to be in the state
$\psi$', while ${T_\phi}^{\!\!\ast}$ describes the process of
`observing the system to be in the state $\phi$'.  Given this,
${T_\phi}^{\!\!\ast} \, T_\psi$ should describe the process of first
preparing the system to be in the state $\psi$ and then observing it
to be in the state $\phi$.  The above equation then relates this
composite process to the transition amplitude $\langle \phi, \psi
\rangle$.  Moreover, we see that `observation' is like a time-reversed
version of `preparation'.  All this makes a rough intuitive kind of
sense.  However, these ideas could use a great deal of elaboration and
clarification.  I mention them here mainly to open an avenue for
further thought.

Second, and less speculatively, the equation $Z(M^\ast) = Z(M)^\ast$
sheds some light on the relation between topology change and the
failure of unitarity, mentioned already in Section \ref{TQFT}.  In any
$\ast$-category, we may define a morphism $f \maps x \to y$ to be {\bf
unitary} if $f^\ast f = 1_x$ and $ff^\ast = 1_y$.  For a morphism in
$\Hilb$ this reduces to the usual definition of unitarity for a linear
operator.  One can show that a morphism $M$ in $n\Cob$ is unitary if 
$M$ involves no topology change, or more precisely, if $M$ is diffeomorphic 
to the Cartesian product of an interval and some $(n-1)$-dimensional manifold.  
(The converse is true in dimensions $n \le 3$, but it fails in higher 
dimensions.)  A TQFT satisfying $Z(M^\ast) = Z(M)^\ast$ maps
unitary morphisms in $n\Cob$ to unitary morphisms in $\Hilb$, so
for TQFTs of this sort, {\it absence of topology change
implies unitary time evolution}.  This fact reinforces a point already 
well-known from quantum field theory on curved spacetime, namely that 
unitary time evolution is not a built-in feature of quantum theory 
but rather the consequence of specific assumptions about the nature 
of spacetime \cite{Earman}.

To conclude, it is interesting to contrast $n\Cob$ and $\Hilb$ with
the more familiar category $\Set$, whose objects are sets and whose
morphisms are functions.  There is no way to make $\Set$ into a
$\ast$-category, since there is no way to `reverse' the map from the
empty set to the one-element set.  So, our intuitions about sets and
functions help us very little in understanding $\ast$-categories.  The
problem is that the concept of function is based on an intuitive
notion of process that is asymmetrical with respect to past and
future: a function $f \maps S \to S'$ is a relation such that each
element of $S$ is related to exactly one element of $S'$, but not
necessarily vice versa.  For better or worse, this built-in `arrow of
time' has no place in the basic concepts of quantum theory.

Pondering this, it soon becomes apparent that if we want an easy 
example of a $\ast$-category other than $\Hilb$ to help build our 
intuitions about $\ast$-categories, we should use not $\Set$ but 
$\Rel$, the category of sets and {\it relations}.  In fact, quantum 
theory can be seen as a modified version of the theory of relations in 
which Boolean algebra has been replaced by the algebra of complex 
numbers!  To see this, note that a linear operator between two Hilbert 
spaces can be described using a matrix of complex numbers as soon as we 
pick an orthonormal basis for each.  Similarly, a relation $R$ between 
sets $S$ and $S'$ can be described by a matrix of truth values, 
namely the truth values of the propositions $yRx$ where $x \in S$
and $y \in S'$.  Composition of relations can be defined as matrix 
multiplication with `or' and `and' playing the roles of `plus' and 
`times'.  It easy to check that this is associative and has an
identity morphism for each set, so we obtain a category $\Rel$ with 
sets as objects and relations as morphisms.  Furthermore, $\Rel$ 
becomes a $\ast$-category if we define the relation $R^\ast$ by saying 
that $xR^\ast y$ if and only if $yRx$.  Just as the matrix for the linear 
operator $T^\ast$ is the conjugate transpose of the matrix for $T$, the 
matrix for the relation $R^\ast$ is the transpose of the matrix for $R$.  

So, the category of Hilbert spaces closely resembles the category of
relations.  The main difference is that binary truth values describing
whether or not a transition is possible are replaced by complex
numbers describing the amplitude with which a transition occurs.
Comparisons between $\Hilb$ and $\Rel$ are fruitful source of
intuitions not only about $\ast$-categories in general but also about
the meaning of `matrix mechanics'.  For some further explorations
along these lines, see the work of Abramsky and Coecke \cite{AC}.

\section{The Monoidal Category of Hilbert Spaces} 
\label{monoidal}

An important goal of the enterprise of physics is to describe, not
just one physical system at a time, but also how a large complicated
system can be built out of smaller simpler ones.  The simplest case is
a so-called `joint system': a system built out of two separate parts.
Our experience with the everyday world leads us to believe that to
specify the state of a joint system, it is necessary and sufficient to
specify states of its two parts.  (Here and in what follows, by
`states' we always mean what physicists call `pure states'.)  In other
words, a state of the joint system is just an ordered pair of states
of its parts.  So, if the first part has $S$ as its set of states, and
the second part has $T$ as its set of states, the joint system has the
cartesian product $S \times T$ as its set of states.
  
One of the more shocking discoveries of the twentieth century is that
this is {\it wrong}.  In both classical and quantum physics, given
states of each part we get a state of the joint system.  But only in
classical physics is every state of the joint system of this form!  In
quantum physics are also `entangled' states, which can only be
described as {\it superpositions} of states of this form.  The reason
is that in quantum theory, the states of a system are no longer
described by a set, but by a Hilbert space.  Moreover --- and this is
really an extra assumption --- the states of a joint system are
described not by the cartesian product of Hilbert spaces, but by their
tensor product.

Quite generally, we can imagine using objects in any category to
describe physical systems, and morphisms between these to describe
processes.  In order to handle joint systems, this category will need
to have some sort of `tensor product' that gives an object $A \tensor
B$ for any pair of objects $A$ and $B$.  As we shall explain,
categories of this sort are called `monoidal'.  The category $\Set$ is
a example where the tensor product is just the usual cartesian product
of sets.  Similarly, the category $\Hilb$ is a monoidal category where
the tensor product is the usual tensor product of Hilbert spaces.
However, these two examples are very different, because the product in
$\Set$ is `cartesian' in a certain technical sense, while the product
in $\Hilb$ is not.  This turns out to explain a lot about why joint
systems behave so counterintuively in quantum physics.  Moreover, it
is yet another way in which $\Hilb$ resembles $n\Cob$ more than
$\Set$.

To see this in detail, it pays to go back to the beginning and think
about cartesian products.  Given two sets $S$ and $T$, we define $S
\times T$ to be the set of all ordered pairs $(s,t)$ with $s \in S$
and $t \in T$.  But what is an ordered pair?  This depends on our
approach to set theory.  We can use axioms in which ordered pairs are
a primitive construction, or we can define them in terms of other
concepts.  For example, in 1914, Wiener defined the ordered pair
$(s,t)$ to be the set $\{ \{ \{s\}, \emptyset \}, \{ \{t\} \}$.  In
1922, Kuratowski gave the simpler definition $(s,t) =
\{ \{s\}, \{s,t\} \}$.  We can use the still simpler definition 
$(s,t) = \{s,\{s,t\}\}$ if our axioms exclude the possibility of sets
that contain themselves.  Various other definitions have also been
tried \cite{Kanamori}.  In traditional set theory we arbitrarily
choose one approach to ordered pairs and then stick with it.  Apart
from issues of convenience or elegance, it does not matter which we
choose, so long as it `gets the job done'.  In other words, all these
approaches are all just technical tricks for implementing our goal, which
is to make sure that $(s,t) = (s',t')$ if and only if $s = s'$ and $t = t'$.

It is a bit annoying that the definition of ordered pair cannot get
straight to the point and capture the concept without recourse to an
arbitrary trick.  It is natural to seek an approach that focuses more
on the {\it structural role} of ordered pairs in mathematics and less
on their {\it implementation}.  This is what category theory provides.

The reason traditional set theory arbitarily chooses a specific
implementation of the ordered pair concept is that it seems difficult
to speak precisely about ``some thing $(s,t)$ --- I don't care what it
is --- with the property that $(s,t) = (s',t')$ iff $s = s'$ and $t =
t'$''.  So, the first move in category theory is to stop focussing on
ordered pairs and instead focus on cartesian products of sets.  What
properties should the cartesian product $S \times T$ have?  To make
our answer applicable not just to sets but to objects of other
categories, it should not refer to elements of $S \times T$.  So, the
second move in category theory is to describe the cartesian product $S
\times T$ in terms of functions to and from this set.

The cartesian product $S \times T$ has 
functions called `projections' to the sets $S$ and $T$:
\[           p_1 \maps S \times T \to S , \qquad
           p_2 \maps S \times T \to T .\]
Secretly we know that these pick out the first or second 
component of any ordered pair in $S \times T$:
\[          p_1(s,t) = s, \qquad p_2(s,t) = t  .\]
But, our goal is to characterize the product by means of these
projections without explicit reference to ordered pairs.  For this,
the key property of the projections is that given any element $s \in
S$ and any element $t \in T$, there exists a unique element $x \in S
\times T$ such that $p_1(x) = s$ and $p_2(x) = T$.  Furthermore, as a
substitute for elements of the sets $S$ and $T$, we can use functions
from an arbitrary set to these sets.

Thus, given two sets $S$ and $T$, we define their {\bf 
cartesian product} to be any set $S \times T$ equipped with 
functions $p_1 \maps S \times T \to S$, $p_2 \maps S \times T \to T$
such that for any set $X$ and functions $f_1 \maps X \to S$, 
$f_2 \maps X \to T$, there exists a unique function 
$f \maps X \to S \times T$ with 
\[        f_1 = p_1 f, \qquad  f_2 = p_2 f.  \]
Note that with this definition, the cartesian product is not unique!
Wiener's definition of ordered pairs gives a cartesian product of the
sets $S$ and $T$, but so does Kuratowski's, and so does any
other definition that `gets the job done'.  However, this does not
lead to any confusion, since one can easily show that any two choices
of cartesian product are isomorphic in a canonical way.  For a proof
of this and other facts about cartesian products, see for example the
textbook by McLarty \cite{McLarty}.

All this generalizes painlessly to an arbitrary category.
Given two objects $A$ and $B$ in some category, 
we define their {\bf cartesian product} (or simply {\bf product})
to be any object $A \times B$ equipped with morphisms 
\[          p_1 \maps A \times B \to A, \qquad
          p_2 \maps A \times B \to B,  \]
called {\bf projections}, such that for any object $X$ and morphisms
$f_1 \maps X \to A$, $f_2 \maps X \to B$, there is a unique morphism
$f \maps X \to A \times B$ with $f_1 = p_1 f$ and $f_2 = p_2 f$.  The
product may not exist, and it may not be unique, but it is unique up
to a canonical isomorphism.  Category theorists therefore feel free to
speak of `the' product when it exists.

We say a category has {\bf binary products} if every pair of objects
has a a product.  One can also talk about $n$-ary products for other
values of $n$, but a category with binary products has $n$-ary
products for all $n \ge 1$, since we can construct these as iterated
binary products.  The case $n = 1$ is trivial, since the product of
one object is just that object itself (up to canonical isomorphism).
The only remaining case is $n = 0$.  This is surprisingly important.
A $0$-ary product is usually called a {\bf terminal object} and
denoted $1$: it is an object such that that for any object $X$ there
exists a unique morphism from $X$ to $1$.  Terminal objects are unique
up to canonical isomorphism, so we feel free to speak of `the'
terminal object in a category when one exists.  The reason we denote
the terminal object by $1$ is that in $\Set$, any set with one element
is a terminal object.  If a category has a terminal object and binary
products, it has $n$-ary products for all $n$, so we say it has {\bf
finite products}.

It turns out that these concepts capture much of our intuition about
joint systems in classical physics.  In the most stripped-down version
of classical physics, the states of a system are described as elements
of a mere {\it set}.  In more elaborate versions, the states of a
system form an object in some fancier category, such as the category
of {\it topological spaces} or {\it manifolds}.  But, just like
$\Set$, these fancier categories have finite products --- and we use
this fact when describing the states of a joint system.

To sketch how this works in general, suppose we have any category with
finite products.  To do physics with this, we think of any of the
objects of this category as describing some physical system.  It
sounds a bit vague to say that a physical system is `described by'
some object $A$, but we can make this more precise by saying that
states of this system are morphisms $f \maps 1 \to A$.  When our
category is $\Set$, a morphism of this sort simply picks out an
element of the set $A$.  In the category of topological spaces, a
morphism of this sort picks out a point in the topological space $A$
--- and similarly for the category of manifolds, and so on.  For this
reason, category theorists call a morphism $f \maps 1 \to A$ an {\bf
element} of the object $A$.  

Next, we think of any morphism $g \maps A \to B$ as a `process' carrying 
states of the system described by $A$ to states of the system described 
by $B$.  This works as follows: given a state of the first system, 
say $f \maps 1 \to A$, we can compose it with $g$ to get a state of
the second system, $gf \maps 1 \to B$.

Then, given two systems that are described by the objects $A$ and $B$,
respectively, we decree that the joint system built from these is
described by the object $A \times B$.  The projection $p_1 \maps A
\times B \to A$ can be thought of as a process that takes a state of
the joint system and discards all information about the second part,
retaining only the state of the first part.  Similarly, the projection
$p_2$ retains only information about the second part.  

Calling these projections `processes' may strike the reader as
strange, since `discarding information' sounds like a subjective
change of our {\it description} of the system, rather than an
objective physical process like time evolution.  However, it is worth
noting that in special relativity, time evolution corresponds to a
change of coordinates $t \mapsto t + c$, which can also be thought of
as change of our description of the system.  The novelty in thinking
of a projection as a physical process really comes, not from the fact
that it is `subjective', but from the fact that it is not invertible.

With this groundwork laid, we can use the definition of `product' to
show that a state of a joint system is just an ordered pair of states
of each part.  First suppose we have states of each part, say $f_1
\maps 1 \to A$ and $f_2 \maps 1 \to B$.  Then there is a unique state
of the joint system, say $f \maps 1 \to A \times B$, which reduces to
the given state of each part when we discard information about the
other part: $p_1 f = f_1$ and $p_2 f = f_2$.  Conversely, every state
of the joint system arises this way, since given $f \maps 1 \to A
\times B$ we can recover $f_1$ and $f_2$ using these equations.

However, the situation changes drastically when we switch to 
quantum theory!  The states of a quantum system can still be 
thought of as forming a set.  However, we do not 
take the product of these sets to be the set of 
states for a joint quantum system.   Instead, we describe 
states of a system as unit vectors in a Hilbert space, modulo
phase.  We define the Hilbert space for a joint system to be the
{\it tensor product} of the Hilbert spaces for its parts.

The tensor product of Hilbert spaces is not a cartesian product
in the sense defined above, since given Hilbert spaces
$H$ and $K$ there are no linear operators
$p_1 \maps H \tensor K \to H$ and $p_2 \maps H \tensor K \to K$
with the required properties.  This means that from a
(pure) state of a joint quantum system we cannot extract 
(pure) states of its parts.  This is the key to Bell's `failure
of local realism'.  Indeed, under quite general conditions
one can derive Bell's inequality from the assumption that pure
states of a joint system determine pure states
of its parts \cite{B0,Bell}, so violations of Bell's inequality should
be seen as an indication that this assumption fails.  

The Wooters--Zurek argument that `one cannot clone a quantum state' 
\cite{WZ} is also based on the fact that the tensor product of Hilbert 
spaces is not cartesian.  To get some sense of this, note that 
whenever $A$ is an object in some category for which the product 
$A \times A$ exists, there is a unique morphism 
\[ \Delta \maps A \to A \times A \]
such that $p_1 \Delta = 1_A$ and $p_2 \Delta = 1_A$.  This morphism is
called the {\bf diagonal} of $A$, since in the category of sets it
is the map given by $\Delta(a) = (a,a)$ for all $a \in A$, whose graph
is a diagonal line when $A$ is the set of real numbers.
Conceptually, the role of a diagonal morphism is to {\it duplicate}
information, just as the projections {\it discard} information.  In
applications to physics, the equations $p_1 \Delta = 1_A$ and 
$p_2 \Delta = 1_A$ says that if we duplicate a state in $A$ and then 
discard one of the two resulting copies, we are left with a copy 
identical to the original.  

In $\Hilb$, however, since the tensor product is not a product in the
category-theoretic sense, it makes no sense to speak of a diagonal
morphism $\Delta \maps H \to H \tensor H$.  In fact, a stronger
statement is true: there is no natural (i.e.\ basis-independent) way
to choose a linear operator from $H$ to $H \tensor H$ other than the
zero operator.  So, there is no way to duplicate information in
quantum theory.

Since the tensor product is not a cartesian product in the sense
explained above, what exactly is it?  To answer this, we need the
definition of a `monoidal category'.  Monoidal categories were
introduced by Mac Lane \cite{MacLane} in early 1960s, precisely in
order to capture those features common to all categories equipped with
a well-behaved but not necessarily cartesian product.  Since the
definition is a bit long, let us first present it and then discuss it:

\vskip 1em
\noindent 
{\bf Definition. }  {\it A {\bf monoidal category} consists of:
\begin{description}
    \item[(i)] a category $\M$,
    \item[(ii)] a functor $\tensor \maps \M \times \M \to \M$,
    \item[(iii)] a {\bf unit object} $I \in \M$,
    \item[(iv)] natural isomorphisms called the {\bf associator}:
     \[ a_{A,B,C} \maps (A \ten B) \ten C \to A \ten (B \ten C), \]
     the {\bf left unit law}:
     \[ \ell_A \maps I \ten A \to A , \]
     and the {\bf right unit law}:
     \[ r_A \maps A \ten I \to A,  \]
\end{description}
such that the following diagrams commute for all objects $A,B,C,D \in 
\M$:
\begin{description}
    \item[(v)]
\[
  \xy 0 ;/r.30pc/:
    (0,20)*{(A \ten B)\ten (C \ten D)}="1";
    (40,0)*{A \ten (B \ten (C \ten D))}="2";
    (25,-20)*{ \quad A \ten ((B \ten C) \ten D)}="3";
    (-25,-20)*{(A \ten (B \ten C)) \ten D}="4";
    (-40,0)*{((A \ten B) \ten C) \ten D}="5";
        {\ar^{a_{A,B,C \ten D}}     "1";"2"}
        {\ar_{1_A \ten a_{B,C,D}}  "3";"2"}
        {\ar^{a_{A,B \ten C,D}}    "4";"3"}
        {\ar_{a_{A,B,C} \ten 1_D}  "5";"4"}
        {\ar^{a_{A \ten B,C,D}}    "5";"1"}
 \endxy
 \\
\]
\vskip 1em
    \item[(vi)]
\[
 \xymatrix{
    (A \ten I) \ten B        \ar[rr]^{a_{A,I,B}}
        \ar[dr]_{r_A \ten 1_B}
     &&  A \ten (I \ten B)
        \ar[dl]^{1_A \ten \ell_B } \\
    & A \ten B   } \\
\]
\end{description}
}

\vskip 1em

This obviously requires some explanation!  First, it makes use of some
notions we have not explained yet, ruining our otherwise admirably
self-contained treatment of category theory.  For example, what is $\M
\times \M$ in clause (ii) of the definition?  This is just the
category whose objects are pairs of objects in $\M$, and whose
morphisms are pairs of morphisms in $\M$, with composition of
morphisms done componentwise.  So, when we say that the tensor product
is a functor $\tensor \maps \M \times \M \to \M$, this implies
that for any pair of objects $x,y \in \M$ there is an object
$x \tensor y \in \M$, while for any pair of morphisms $f \maps x \to
x', g \maps y \to y'$ in $\M$ there is a morphism $f \tensor g \maps x
\tensor y \to x' \tensor y'$ in $\M$.  Morphisms are just as important
as objects!  For example, in $\Hilb$, not only can we take the tensor
product of Hilbert spaces, but also we can take the tensor product of
bounded linear operators $S \maps H \to H'$ and $T \maps K \to K'$, 
obtaining a bounded linear operator
\[  S \tensor T \maps H \tensor K \to H' \tensor K'.  \]
In physics, we think of $S \tensor T$ as a joint process built from
the processes $S$ and $T$ `running in parallel'.  For example, if we
have a joint quantum system whose two parts evolve in time without
interacting, any time evolution operator for the whole system is
given by the tensor product of time evolution operators for the
two parts.

\begin{figure}[h]
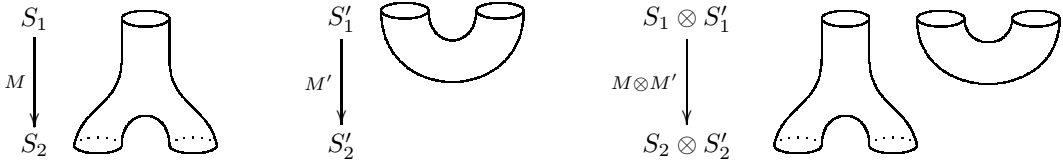

\vskip 2em
\[
 \xy 0;/r.25pc/:
  (0,4)*\ellipse(3,1){-};
  (-3,-4)*\ellipse(3,1){.};
  (-3,-4)*\ellipse(3,1)__,=:a(-180){-};
  (3,-4)*\ellipse(3,1){.};
  (3,-4)*\ellipse(3,1)__,=:a(-180){-};
  (-3,-8)*{}="1";
  (3,-8)*{}="2";
  (-9,-8)*{}="A2";
  (9,-8)*{}="B2";
    "1";"2" **\crv{(-3,-3) & (3,-3)};
  (-3,8)*{}="A";
  (3,8)*{}="B";
  (-3,3)*{}="A1";
  (3,3)*{}="B1";
   "A";"A1" **\dir{-};
   "B";"B1" **\dir{-};
    "B2";"B1" **\crv{(8,-3) & (3,-3)};
    "A2";"A1" **\crv{(-8,-3) & (-3,-3)};
    {\ar_{M} (-14,8)*+{S_1}; (-14,-8)*+{S_2}}
 \endxy
\qquad \quad
  \xy 0;/r.25pc/:
  (3,4.5)*\ellipse(3,1){-};
  (-3,4.5)*\ellipse(3,1){-};
  (-3,9)*{}="X1";
  (3,9)*{}="X2";
    "X1";"X2" **\crv{(-3,4) & (3,4)};
  (-9,9)*{}="X1";
  (9,9)*{}="X2";
    "X1";"X2" **\crv{(-9,-3) & (9,-3)};
    {\ar_{M'} (-14,8)*+{S'_1}; (-14,-8)*+{S'_2}}
 \endxy
\qquad \quad
 \xy 0;/r.25pc/:
  (0,4)*\ellipse(3,1){-};
  (-3,-4)*\ellipse(3,1){.};
  (-3,-4)*\ellipse(3,1)__,=:a(-180){-};
  (3,-4)*\ellipse(3,1){.};
  (3,-4)*\ellipse(3,1)__,=:a(-180){-};
  (-3,-8)*{}="1";
  (3,-8)*{}="2";
  (-9,-8)*{}="A2";
  (9,-8)*{}="B2";
    "1";"2" **\crv{(-3,-3) & (3,-3)};
  (-3,8)*{}="A";
  (3,8)*{}="B";
  (-3,3)*{}="A1";
  (3,3)*{}="B1";
   "A";"A1" **\dir{-};
   "B";"B1" **\dir{-};
    "B2";"B1" **\crv{(8,-3) & (3,-3)};
    "A2";"A1" **\crv{(-8,-3) & (-3,-3)};
    {\ar_{M \tensor M'} (-20,8)*+{S_1 \tensor S'_1}; (-20,-8)*+{S_2 \tensor S'_2}};
  (12,4)*\ellipse(3,1){-};
  (6,4)*\ellipse(3,1){-};
  (15,8)*{}="X1";
  (21,8)*{}="X2";
    "X1";"X2" **\crv{(15,4) & (21,4)};
  (9,8)*{}="X1";
  (27,8)*{}="X2";
    "X1";"X2" **\crv{(9,-3) & (27,-3)};
 \endxy
\]
\medskip
\caption{Two cobordisms and their tensor product}
\label{tensor}
\end{figure}

Similarly, in $n\Cob$ the tensor product is given by disjoint union,
both for objects and for morphisms.  In Figure \ref{tensor} we show
two spacetimes $M$ and $M'$ and their tensor product $M \tensor M'$.
This as a way of letting two spacetimes `run in parallel', like
independently evolving separate universes.  The resemblance to the
tensor product of morphisms in $\Hilb$ should be clear.  
Just as in $\Hilb$, the tensor product in $n\Cob$ is not a cartesian
product: there are no projections with the required properties.  There
is also no natural choice of a cobordism from $S$ to $S \tensor S$.
This means that the very nature of topology prevents us from finding
spacetimes that `discard' part of space, or `duplicate' space.  Seen
in this light, the fact that we cannot discard or duplicate
information in quantum theory is not a flaw or peculiarity of this
theory.  {\it It is a further reflection of the deep structural
analogy between quantum theory and the conception of spacetime
embodied in general relativity.}

Turning to clause (iii) in the definition, we see that a monoidal
category needs to have a `unit object' $I$.  This serves as the
multiplicative identity for the tensor product, at least up to
isomorphism: as we shall see in the next clause, $I \tensor A \cong A$
and $A \tensor I \iso A$ for every object $A \in \M$.  In $\Hilb$ the
unit object is $\C$ regarded as a Hilbert space, while in $n\Cob$ it
is the empty set regarded as an $(n-1)$-dimensional manifold.  Any
category with finite products gives a monoidal category in which the
unit object is the terminal object $1$.

This raises an interesting point of comparison.  In classical physics 
we describe systems using objects in a category with finite products, 
and a state of the system corresponding to the object $A$ is just a 
morphism $f \maps 1 \to A$.  In quantum physics we describe systems 
using Hilbert spaces.  Is a state of the system corresponding to the 
Hilbert space $H$ the same as a bounded linear operator $T \maps
\C \to H$?  Almost, but not quite!  As we saw in Section \ref{*category},
such operators are in one-to-one correspondence with vectors in $H$: 
any vector $\psi \in H$ corresponds to an operator $T_\psi \maps \C \to H$ 
with $T_\psi(1) = \psi$.  States, on the other hand, are the same as
unit vectors modulo phase.  Any nonzero vector in $H$ gives a state
after we normalize it, but different vectors can give the same state,
and the zero vector does not give a state at all.  So, quantum physics 
is really different from classical physics in this way: we cannot 
define states as morphisms from the unit object.  Nonetheless, we
have seen that the morphisms $T \maps \C \to H$ play a fundamental 
role in quantum theory: they are just Dirac's `kets'.

Next, let us ponder clause (iv) of the definition of monoidal
category.  Here we see that the tensor product is associative,
but only {\it up to a specified isomorphism}, called the `associator'.
For example, in $\Hilb$ we do not have 
$(H \tensor K) \tensor L = H \tensor (K \tensor L)$, 
but there is an obvious isomorphism
\[         a_{H,K,L} \maps (H \tensor K) \tensor L \to 
H \tensor (K \tensor L) \]
given by
\[          a_{H,K,L} ((\psi \tensor \phi) \tensor \eta) = 
                        \psi \tensor (\phi \tensor \eta) .\]
Similarly, we do not have $\C \tensor H = H$ and $H \tensor \C = H$,
but there are obvious isomorphisms
\[        \ell_H \maps \C \tensor H \to H, \qquad
          r_H \maps H \tensor \C \to H  .\]
Moreover, all these isomorphisms are `natural' in a precise sense.
For example, when we say the associator is natural, we mean that
for any bounded linear operators $S \maps H \to H'$, $T \maps K \to K'$,
$U \maps L \to L'$ the following square diagram commutes:
\[
\xymatrix{
 (H \tensor K) \tensor L
  \ar[rr]^{a_{H,K,L}}
  \ar[d]_{(S \tensor T) \tensor U}
&&  H \tensor (K \tensor L)
   \ar[d]^{S \tensor (T \tensor U)}     \\
 (H' \tensor K') \tensor L'
   \ar[rr]^{a_{H',K',L'}}
&&  H' \tensor (K' \tensor L') }
\]
In other words, composing the top morphism with the right-hand one
gives the same result as composing the left-hand one with the bottom
one.  This compatibility condition expresses the fact that no arbitrary
choices are required to define the associator: in particular, it
is defined in a basis-independent manner.  Similar but simpler 
`naturality squares' must commute for the left and right unit laws.  

Finally, what about clauses (v) and (vi) in the definition of monoidal
category?  These are so-called `coherence laws', which let us
manipulate isomorphisms with the same ease as if they were equations.
Repeated use of the associator lets us construct an isomorphism from
any parenthesization of a tensor product of objects to any other
parenthesization --- for example, from $((A \tensor B) \tensor C)
\tensor D$ to $A \tensor (B \tensor (C \tensor D))$.  However, we can
actually construct {\it many} such isomorphisms --- and in this
example, the pentagonal diagram in clause (v) shows two.  We would
like to be sure that all such isomorphisms from one parenthesization
to another are equal.  In his fundamental paper on monoidal categories, 
Mac Lane \cite{MacLane} showed that the commuting pentagon in clause 
(v) guarantees this, not just for a tensor product of four objects,
but for arbitrarily many.  He also showed that clause (vi) gives a
similar guarantee for isomorphisms constructed using the left and
right unit laws.

\section{Conclusions} 
\label{conclusions}

Our basic intuitions about mathematics are to some extent
abstracted from our dealings with the everyday physical 
world \cite{Lakoff}.  The concept of a {\it set}, 
for example, formalizes some of our intuitions about piles of pebbles,
herds of sheep and the like.   These things are all pretty
well described by classical physics, at least in their gross
features.  For this reason, it may seem amazing that mathematics based
on set theory can successfully describe the microworld, where
quantum physics reigns supreme.  However, beyond the overall
`surprising effectiveness of mathematics', this should not really
come as a shock.  After all, set theory is sufficiently flexible
that any sort of effective algorithm for making predictions can be 
encoded in the language of set theory: even Peano arithmetic
would suffice.

But, we should not be lulled into accepting the primacy of the
category of sets and functions just because of its flexibility.
The mere fact that we {\it can} use set theory as a 
framework for studying quantum phenomena does not imply that this is 
the most enlightening approach.  Indeed, the famously counter-intuitive
behavior of the microworld suggests that not only
set theory but even classical logic is not optimized for understanding
quantum systems.   While there are no real paradoxes, and one
can compute everything to one's heart's content, one often feels
that one is grasping these systems `indirectly', like a nuclear 
power plant operator handling radioactive material behind a plate glass 
window with robot arms.  This sense of distance is reflected in
the endless literature on `interpretations of quantum mechanics',
and also in the constant invocation of the split between `observer'
and `system'.  It is as if classical logic continued to apply 
to us, while the mysterious rules of quantum theory apply only 
to the physical systems we are studying.  But of course this is not
true: we are part of the world being studied.  

To the category theorist, this raises the possibility that
quantum theory might make more sense when viewed, not from
the category of sets and functions, but {\sl within some other category:} 
for example $\Hilb$, the category of Hilbert spaces and bounded 
linear operators.  Of course it is most convenient to define this 
category and study it with the help of set theory.  However, as we
have seen, the fact that Hilbert spaces are sets equipped with extra
structure and properties is almost a distraction when trying to 
understand $\Hilb$, because its morphisms are not functions that
preserve this extra structure.   So, we can gain a new understanding
of quantum theory by trying to accept $\Hilb$ on its own terms, 
unfettered by preconceptions taken from the category $\Set$.  As 
Corfield \cite{Corfield} writes: ``Category theory allows you to work 
on structures without the need first to pulverise them into set 
theoretic dust.  To give an example from the field of architecture, 
when studying Notre Dame cathedral in Paris, you try to understand how 
the building relates to other cathedrals of the day, and then to earlier 
and later cathedrals, and other kinds of ecclesiastical building.  
What you don't do is begin by imagining it reduced to a pile
of mineral fragments.''

In this paper, we have tried to say quite precisely how some intuitions 
taken from $\Set$ fail in $\Hilb$.   Namely: unlike $\Set$, $\Hilb$ 
is a $\ast$-category, and a monoidal category where the tensor product
is noncartesian.  But, what makes this really interesting is that these 
ways in which $\Hilb$ differs from $\Set$ are precisely the ways 
it resembles $n\Cob$, the category of $(n-1)$-dimensional 
manifolds and $n$-dimensional cobordisms going between these
manifolds.  In general relativity these cobordisms represent 
`spacetimes'.  Thus, from the category-theoretic perspective, a 
bounded linear operator between Hilbert spaces acts more like a 
{\it spacetime} than a {\it function}.  This not only sheds a new light 
on some classic quantum quandaries, it also bodes well for the main 
task of quantum gravity, namely to reconcile quantum theory with general
relativity.

At best, we have only succeeded in sketching a few aspects of the
analogy between $\Hilb$ and $n\Cob$.  In
a more detailed treatment we would explain how both $\Hilb$ and
$n\Cob$ are `symmetric monoidal categories with duals' --- a notion
which subsumes being a monoidal category and a $\ast$-category.
Moreover, we would explain how unitary topological quantum field
theories exploit this fact to this hilt.  However, a discussion of
this can be found elsewhere \cite{BD}, and it necessarily leads
us into deeper mathematical waters which are not of such immediate
philosophical interest.  So, instead, I would like to conclude by
saying a bit about the progress people have made in learning to
think within categories other than $\Set$.

It has been known for quite some time in category theory that
each category has its own `internal logic', and that while we can 
reason externally about a category using classical logic, we can also 
reason {\it within it} using its internal logic --- which gives a very
different perspective.  For example, our best understanding of 
intuitionistic logic has long come from the study of categories called 
`topoi', for which the internal logic differs from classical
logic mainly in its renunciation of the principle 
of excluded middle \cite{BJ,Coste,Mitchell}.  
Other classes of categories have their own forms of internal
logic.  For example, ever since the work of Lambek \cite{Lambek},
the typed lambda-calculus, so beloved by theoretical
computer scientists, has been understood to arise 
as the internal logic of `cartesian closed' categories. 
More generally, Lawvere's algebraic semantics allows
us to see any `algebraic theory' as the internal logic of 
a category with finite products \cite{Lawvere}.  

By now there are many textbook treatments of these ideas and
their ramifications, ranging from introductions that do not assume
prior knowledge of category theory \cite{Crole,McLarty},
to more advanced texts that do \cite{BW,Johnstone,LS,MM}.
Lawvere has also described how to do classical physics in a topos
\cite{Lawvere2,Lawvere3}.
All this suggests that the time is ripe to try thinking about
quantum physics using the internal logic of $\Hilb$, or $n\Cob$, or
related categories.  However, the textbook treatments and even
most of the research literature on category-theoretic logic
focus on categories where the monoidal structure is cartesian.  
The study of logic within more general monoidal categories is 
just beginning.  More precisely, while generalizations of
`algebraic theories' to categories of this sort have been studied 
for many years in topology and physics \cite{LSV,MSS}, it is 
hard to find work that explicitly recognizes the relation of such 
theories to the traditional concerns of logic, or even of quantum logic.
For some heartening counterexamples, see the work of Abramsky
and Coecke \cite{AC}, and also of Mauri \cite{Mauri}.  So, we can 
only hope that in the future, more interaction between mathematics,
physics, logic and philosophy will lead to new ways of thinking about 
quantum theory --- and quantum gravity --- that take advantage of 
the internal logic of categories like $\Hilb$ and $n\Cob$.

\subsection*{Acknowledgements}

I thank Aaron Lauda for preparing most of the figures in this
paper, and thank Bob Coecke, David Corfield, Daniel Ruberman,
Issar Stubbe and especially James Dolan for helpful conversations.  
I would also like to thank the Physics Department of U.\ C.\ Santa Barbara 
for inviting me to speak on this subject, and Fernando Souza
for inviting me to speak on this subject at the American
Mathematical Society meeting at San Francisco State University,
both in October 2000.

\end{document}